\documentclass[aps,prb,reprint,superscriptaddress,longbibliography,floatfix]{revtex4-2}

\usepackage{amsmath,amssymb}
\usepackage{booktabs}
\usepackage{graphicx}
\usepackage{xcolor}
\usepackage{siunitx}
\usepackage{hyperref}
\newcommand{\EviDTB}{EviDTB}

\begin{document}

\title{Evidential Deep Regression for Universal Density Functional Theory Hamiltonian Models}

\author{Jiaxuan Liu}
\affiliation{Beijing National Laboratory for Condensed Matter Physics and Institute of Physics, Chinese Academy of Sciences, Beijing 100190, China}
\affiliation{University of Chinese Academy of Sciences, Beijing 100049, China}

\author{Yan Zhang}
\affiliation{Beijing National Laboratory for Condensed Matter Physics and Institute of Physics, Chinese Academy of Sciences, Beijing 100190, China}
\affiliation{University of Chinese Academy of Sciences, Beijing 100049, China}

\author{Caiyuan Ye}
\affiliation{Beijing National Laboratory for Condensed Matter Physics and Institute of Physics, Chinese Academy of Sciences, Beijing 100190, China}
\affiliation{University of Chinese Academy of Sciences, Beijing 100049, China}

\author{Tiannian Zhu}
\affiliation{Beijing National Laboratory for Condensed Matter Physics and Institute of Physics, Chinese Academy of Sciences, Beijing 100190, China}
\affiliation{University of Chinese Academy of Sciences, Beijing 100049, China}

\author{Zhong Fang}
\affiliation{Beijing National Laboratory for Condensed Matter Physics and Institute of Physics, Chinese Academy of Sciences, Beijing 100190, China}
\affiliation{University of Chinese Academy of Sciences, Beijing 100049, China}

\author{Hongming Weng}
\email{hmweng@iphy.ac.cn}
\affiliation{Beijing National Laboratory for Condensed Matter Physics and Institute of Physics, Chinese Academy of Sciences, Beijing 100190, China}
\affiliation{University of Chinese Academy of Sciences, Beijing 100049, China}

\author{Quansheng Wu}
\email{quansheng.wu@iphy.ac.cn}
\affiliation{Beijing National Laboratory for Condensed Matter Physics and Institute of Physics, Chinese Academy of Sciences, Beijing 100190, China}
\affiliation{University of Chinese Academy of Sciences, Beijing 100049, China}

\date{October 2026}

\begin{abstract}

Universal density functional theory (DFT) Hamiltonian models trained across broad chemical and structural spaces offer an efficient route from atomic structure to electronic properties. However, estimating prediction uncertainty alongside fast Hamiltonian predictions remains a challenge. In this work, we present \EviDTB{}, an evidential deep regression approach for tight-binding DFT Hamiltonians that effectively estimates prediction uncertainty in a single forward pass. Compared with deep ensembles, Monte Carlo dropout, and mean-variance estimation (MVE), \EviDTB{} achieves a better balance between computational cost and uncertainty-quantification performance. On 5,478 held-out test structures, \EviDTB{} achieves a full spin--orbit-coupled (SOC) Hamiltonian matrix-element mean absolute error of \SI{0.622}{\milli\electronvolt}. Its mean uncertainty per atom correlates with the mean local squared prediction error (Spearman $\rho=0.812$); removing the 15\% most uncertain structures reduces the 99th percentile of this error by 78\%, from 16.63 to 3.72~eV$^2$. This score offers a practical way to prioritize candidates for first-principles verification. We hope that combining fast Hamiltonian prediction with uncertainty-guided screening will accelerate the search for materials with desirable electronic properties.
\end{abstract}

\maketitle

\section{Introduction}

Machine learning is playing an increasingly important role in condensed-matter physics and materials science, where the relevant configuration and chemical spaces are often too large to explore solely with conventional numerical calculations~\cite{Carleo2019,Butler2018}.
Its applications now span several stages of materials research.
Machine-learning interatomic potentials reproduce first-principles potential-energy surfaces and enable large-scale atomistic simulations~\cite{Behler2007,Zhang2018DeepPotential,Unke2021}; generative models propose candidate crystals directly in periodic structure space~\cite{Noh2019,Kim2020CrystalGAN,Xie2022CDVAE,Jiao2023DiffCSP,Ye2024ConCDVAE,Zeni2025MatterGen,Cao2025CrystalFormer,Ye2026PODGen}; and symmetry-aware neural networks predict localized quantum operators, including Kohn--Sham Hamiltonians, from which electronic bands and related observables can be obtained. Examples include the DeepH family~\cite{Li2022DeepH,Gong2023DeepHE3,Li2023xDeepH,Wang2024DeepH2}, HamGNN and its charge-aware extension~\cite{Zhong2023HamGNN,Ma2025HamGNNQ}, SLEM as implemented in DeePTB~\cite{Zhouyin2025SLEM,Gu2024DeePTB}, and other equivariant operator-learning approaches~\cite{Unke2021PhiSNet,Yu2023QHNet,Qian2026MACEH}.
Together, these developments are turning machine learning from a task-specific fitting tool into a computational layer that connects structure generation, atomic relaxation, and electronic-property evaluation.

A particularly important recent advance is the emergence of \emph{universal} models trained across broad chemical and structural spaces.
Here, ``universal model'' denotes a model trained across multiple chemical species and structural environments rather than a model specialized to one material family; it does not imply guaranteed reliability for arbitrary out-of-distribution structures.
Universal interatomic potentials such as M3GNet~\cite{Chen2022M3GNet}, CHGNet~\cite{Deng2023CHGNet}, MACE-MP~\cite{Batatia2025MACEMP}, MatterSim~\cite{Yang2024MatterSim}, and the DPA family~\cite{Zhang2022DPA1,Zhang2024DPA2} extend energy and force prediction from a small set of compositions toward substantial portions of the periodic table.
For Hamiltonian prediction, universal DeepH models, Universal HamGNN and its spin--orbit-coupled extension, NextHAM, and the physics-prior-based UniTB have broadened the chemical and structural scope of learned Hamiltonians~\cite{Wang2024UniversalDeepH,Zhong2024UniversalHamiltonian,Zhong2026UniHamGNN,Yin2025NextHAM,Liu2026UniTB}.
By replacing repeated self-consistent calculations with neural-network inference, these models make it possible to relax and electronically characterize far more candidates than would otherwise be practical.
They have consequently become important components of high-throughput materials screening and machine-learning-assisted discovery.

Despite their success on many in-distribution tasks, universal models do not uniformly reproduce the calculation across materials, properties, and conditions. Universal models can fail on configurations not represented in the training data, particularly under extreme conditions~\cite{Subramanyam2025Entropy, Focassio2025Surfaces}. Thus, low average errors alone cannot establish the reliability of a particular material prediction, and targeted first-principles verification remains important.
In practice, machine-learning-assisted high-throughput discovery typically uses rapid screening or structural pre-relaxation to narrow down large candidate pools, followed by DFT calculations to verify the stability and properties of selected materials~\cite{Merchant2023GNoME,Chen2024CloudDiscovery,Choi2026HighEntropy}. Some workflows additionally fine-tune pretrained universal models on targeted first-principles data before using them for property prediction or atomistic simulations~\cite{Lian2025FineTunedNEB,Kim2026MAXPhase,Kong2025MatterTune}.
With a reliable Uncertainty Quantification, it is possible to calculate only a few samples with high uncertainty, banlancing efficiency and accuracy on demand.

A range of uncertainty-quantification strategies has been developed.
Deep ensembles use disagreement among independently trained models and are often reliable, but multiply both training and inference costs~\cite{Lakshminarayanan2017}.
Monte Carlo dropout approximates predictive sampling through repeated stochastic forward passes~\cite{Gal2016}, whereas mean-variance estimation (MVE) learns a conditional variance together with the prediction~\cite{Nix1994}.
Latent-space Gaussian mixture models (GMM) provide a post hoc measure of how atypical a representation is~\cite{Zhu2023GMM}; more recent approaches reduce cost through shared shallow ensembles~\cite{Kellner2024DPOSE} or infer uncertainty from training-loss trajectories~\cite{Vita2025LTAUFF}.
Deep evidential regression instead predicts a distribution over likelihood parameters, allowing aleatoric and epistemic uncertainty to be estimated in a single forward pass~\cite{Amini2020}.

In the field of machine-learning interatomic potentials, uncertainty estimation has already enabled on-the-fly data acquisition in VASP-MLFF~\cite{Jinnouchi2019MLFF}, and several of the aforementioned techniques have been incorporated into deep learning frameworks~\cite{Tan2023UQ,Xu2026EIP}.
These studies demonstrate that these algorithms can effectively identify configurations with higher error while also exposing the tradeoff between reliability and computational overhead.
These advances motivate evaluating efficient uncertainty estimates for universal machine-learning Hamiltonians, whose material-dependent errors directly affect downstream electronic-structure screening.

In this work, we introduce Evidential Deep Tight Binding (EviDTB), built on the DeePTB software~\cite{Gu2024DeePTB}, to adapt evidential deep learning to residual Hamiltonian learning, enabling the reliability of a predicted Hamiltonian to be assessed together with the prediction itself.

\section{Results and Discussion}

\subsection{Framework of EviDTB}

\begin{figure*}[!tp]
 \includegraphics[width=\textwidth]{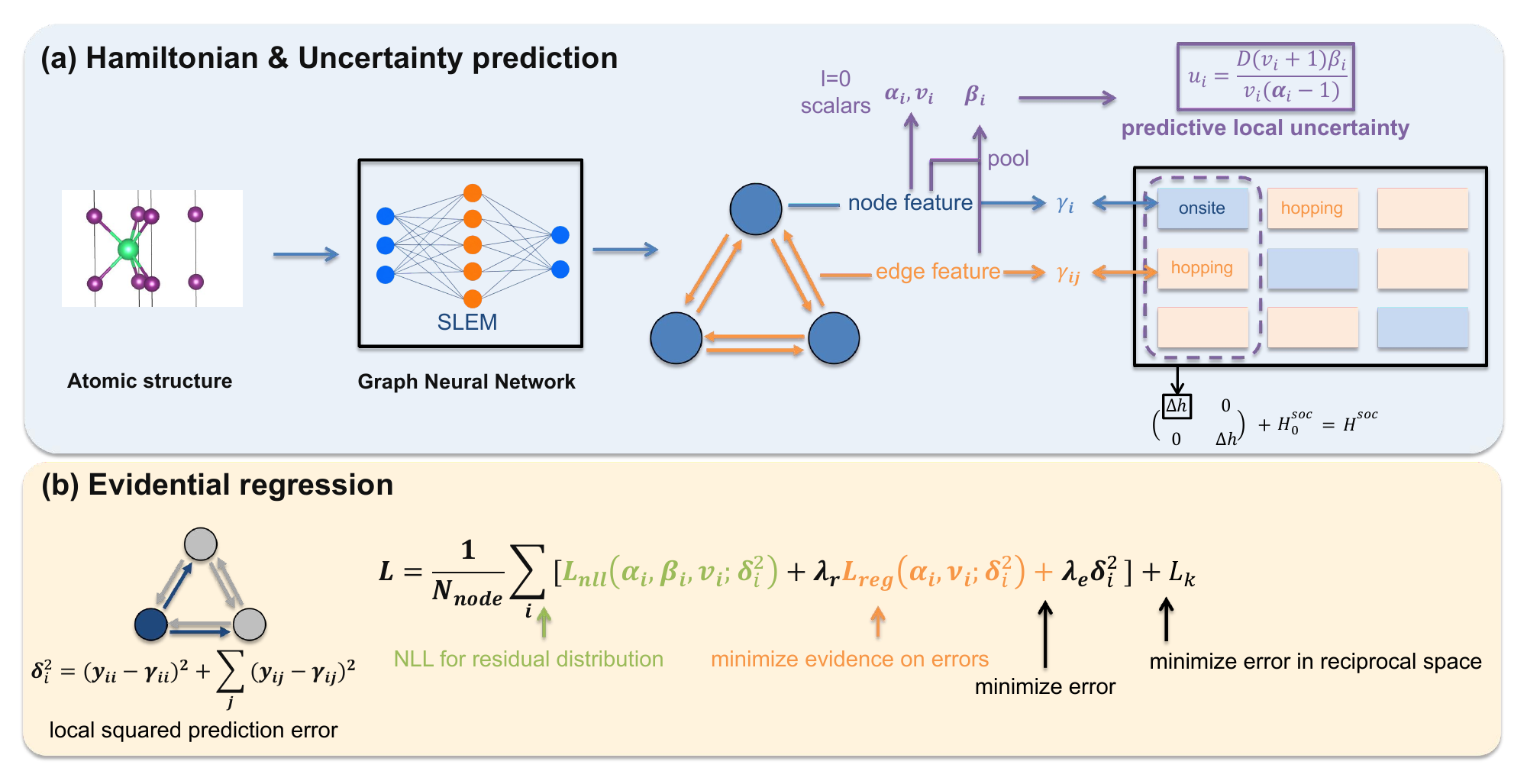}
  \caption{Framework of \EviDTB. (a) SLEM transforms an atomic structure into equivariant node and directed-edge features. Node and edge outputs predict onsite and hopping blocks of the diagonal Hamiltonian residual $\Delta h^{pred}$, respectively. Scalar channels associated with source node $i$ predict $(\alpha_i,\beta_i,\nu_i)$ and give the total predictive uncertainty $u_i$. (b) The source-centered prediction-error norm $\delta_i$ aggregates the onsite error and outgoing hopping errors. Training combines the corresponding evidential negative log-likelihood, an evidence regularizer, a direct real-space squared-error term, and a reciprocal-space loss $L_k$.}
 \label{fig:framework}
\end{figure*}

Our target is the self-consistent nonmagnetic DFT Hamiltonian of a crystal with spin--orbit coupling (SOC), represented in a nonorthogonal basis of localized atomic orbitals.

Self-consistent DFT calculations commonly begin with a superposition of atomic charge densities as the initial charge-density guess. The Hamiltonian evaluated at this initial density provides a structure-dependent reference Hamiltonian, $H_0^{\mathrm{SOC}}$, that can be constructed without performing self-consistent iterations. Following the $\Delta$-learning strategy of NextHAM~\cite{Yin2025NextHAM}, we learn only the residual $H^{\mathrm{SOC}}-H_0^{\mathrm{SOC}}$ between the converged and reference Hamiltonians, rather than the full Hamiltonian.

In OpenMX, SOC is incorporated through fully relativistic, total-angular-momentum-dependent nonlocal pseudopotentials~\cite{Theurich2001SOC}. For a fixed atomic structure and a fixed atomic-orbital basis, this pseudopotential SOC contribution remains unchanged during self-consistent iterations~\cite{Verstraete2008SOC}. In the nonmagnetic PBE calculations considered here, the density-dependent Hartree and exchange--correlation correction is spin independent and has real matrix elements in the real-valued atomic-orbital basis, contributing only to the real parts of the spin-diagonal blocks, up to numerical precision. The converged Hamiltonian can therefore be written in spinor form as

\begin{equation}
 H^{\mathrm{SOC}}=H_0^{\mathrm{SOC}}+
 \begin{pmatrix}\Delta h&0\\0&\Delta h\end{pmatrix},
 \label{eq:soc_residual}
\end{equation}

where the real symmetric matrix $\Delta h$ represents the real-space residual learned by the neural network.

To learn $\Delta h^{pred}$, we adopt SLEM~\cite{Zhouyin2025SLEM}, a strictly localized $E(3)$-equivariant graph neural network implemented in the DeePTB package~\cite{Gu2024DeePTB}, as the backbone of \EviDTB.
As illustrated in Fig.~\ref{fig:framework}(a), SLEM represents atoms as nodes and localized orbital interactions as directed edges, and constructs equivariant node and edge features from the atomic species and geometry.
The node outputs (denoted as $\gamma_i$) describe onsite blocks, whereas the edge outputs (denoted as $\gamma_{ij}$) in both directions describe hopping blocks. The full predicted matrix is then reconstructed by enforcing Hermiticity (symmetry for the real residual considered here).

In parallel, we follow the deep evidential regression framework ~\cite{Amini2020} to estimate the uncertainty of the predicted Hamiltonian. In this framework, the target is assumed to follow a Gaussian likelihood, whose mean $\mu$ and variance $\sigma^2$ are unknown. The model predicts parameters $(\alpha,\beta,\gamma,\nu)$ defining a normal--inverse-gamma (NIG) distribution over $\mu$ and $\sigma^2$. Finally, the prediction and the aleatoric and epistemic variances are extracted from those parameters:

\[
\underbrace{\mathbb{E}[\mu]=\gamma}_{\text{prediction}},
\qquad
\underbrace{\mathbb{E}[\sigma^2]=\frac{\beta}{\alpha-1}}
          _{\text{aleatoric}},
\qquad
\underbrace{\operatorname{Var}[\mu]=\frac{\beta}{\nu(\alpha-1)}}
          _{\text{epistemic}}.
\]

In EviDTB implementation,
we use scalar ($l=0$) channels associated with source node $i$ to predict the normal--inverse-gamma parameters $(\alpha_i,\nu_i)$, while $\beta_i$ is obtained by summing the squared scales predicted from the node and its outgoing-edge features.
Together with the Hamiltonian prediction written in features $\gamma_i,\gamma_{ij}$, these parameters define a source-centered Gaussian--NIG predictive model in an effective low-dimensional subspace.
We assume that local prediction errors and fluctuations of the unknown mean occupy an effective $D$-dimensional subspace, and adopt an isotropic Gaussian--NIG model within that subspace. The resulting total predictive variance is:

\begin{equation}
 u_i=D\left[\frac{\beta_i}{\alpha_i-1}
 +\frac{\beta_i}{\nu_i(\alpha_i-1)}\right]
 =\frac{D\beta_i(1+\nu_i)}{\nu_i(\alpha_i-1)},
 \label{eq:uncertainty}
\end{equation}

The model predicts the parameters $(\alpha_i,\beta_i,\nu_i)$ for each node to estimate uncertainty alongside the Hamiltonian correction in a single forward pass.

Figure~\ref{fig:framework}(b) summarizes how the local evidential training target is constructed.
For each source node $i$, the local squared prediction error $\delta_i^2$ combines the squared error of $\gamma_i$ with the squared errors of $\gamma_{ij}$.

$\delta_i^2$ and the evidential parameters $(\alpha_i,\beta_i,\nu_i)$ define the NLL and evidence-regularization terms used to optimize the evidential parameters. We also include a direct real-space squared-error term and a reciprocal-space loss to improve prediction accuracy. Details are given in Methods.

\subsection{Data preparation}

\begin{figure}[!htbp]
 \includegraphics[width=\columnwidth]{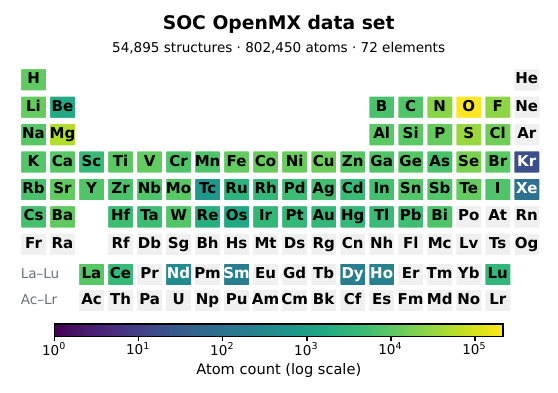}
  \caption{Element distribution of the valid SOC OpenMX data set. Cells show element symbols; color represents the number of atoms of each element on a logarithmic scale. Gray cells denote elements with zero occurrence in the valid data set.}
 \label{fig:dataset_elements}
\end{figure}

\begin{figure*}[!tp]
 \includegraphics[width=\textwidth]{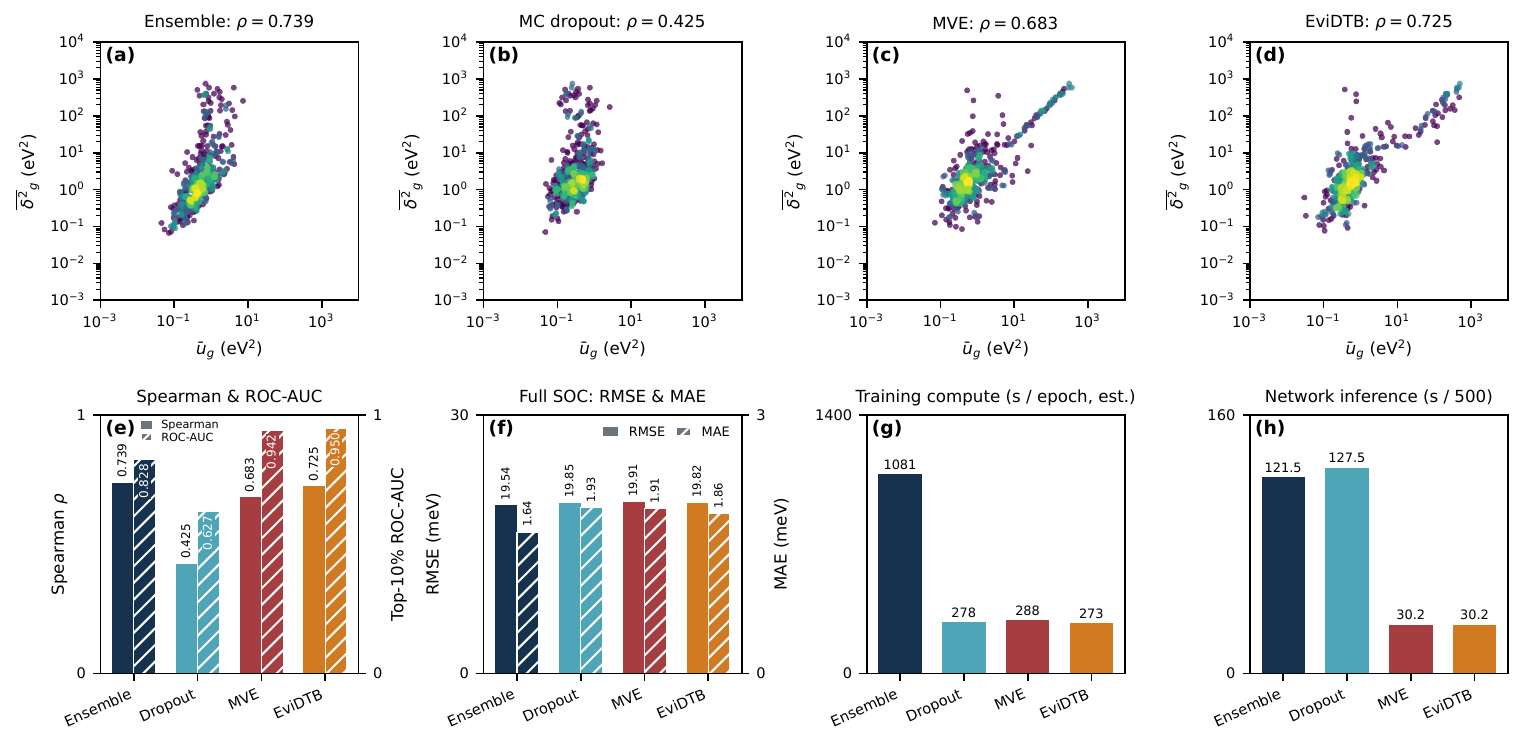}
   \caption{Uncertainty and computational-cost comparison on the test split of the small subset. (a)--(d) Mean uncertainty per atom, $\bar u_g$, versus mean local squared prediction error, $\overline{\delta^2}_g$, for the four methods, plotted with identical log--log axes and density coloring. Both quantities are in eV$^2$. (e) Spearman rank correlation (left axis, solid bars) and ROC-AUC for detecting each method's highest-error 10\% of test structures (right axis, hatched bars), using that method's mean local squared prediction error to define the positive class. Statistics use all 500 structures in the test set. (f) Matrix-element root mean squared error (RMSE) and mean absolute error (MAE) of the reconstructed full complex SOC Hamiltonians for all four methods. (g) Estimated training-compute time per epoch per rank; one ensemble epoch includes all four members. (h) Measured inference-compute time for all 500 test structures; the dropout timing uses 4 stochastic passes.}
 \label{fig:method_comparison}
\end{figure*}

We obtained crystal structures from the Materials Project~\cite{Jain2013MaterialsProject} and prepared 58,685 OpenMX calculation candidates containing no more than 40 atoms each~\cite{Ozaki2003OpenMX}. The calculations used the PBE exchange--correlation functional and element-dependent OpenMX PBE19-family pseudopotentials in a localized pseudoatomic-orbital basis. Retaining successfully completed calculations  yielded 54,895 structures for the final data set. These structures contain 802,450 atoms from 72 elements; their elemental distribution is shown in Fig.~\ref{fig:dataset_elements}. The training, validation, and held-out test splits comprise 43,915, 5,502, and 5,478 structures, respectively (Table~\ref{tab:protocols}).

\begin{table}[!htbp]
\caption{Data sets and evaluation splits. }
\label{tab:protocols}
\scriptsize
\begin{tabular}{l c c c}
Data set & Training & Validation & Test  \\
\hline
Small subset & 4,000 & 500 & 500  \\
Full data set & 43,915 & 5,502 & 5,478  \\
Generated-structure test set & -- & -- & 636  \\
\end{tabular}
\end{table}

The spin-block norms supporting the real spin-diagonal correction are reported in Sec.~S3 and Table~S1 of the Supplementary Information.

\subsection{Experiments}

We first compare uncertainty estimators on a subset of 5,000 structures with OpenMX SOC Hamiltonians. We then train EviDTB on the full data set, using 43,915 training and 5,502 validation structures, and evaluate it on 5,478 held-out structures and an external generated-structure set.

\subsubsection{Uncertainty-estimator comparison.}

We compare EviDTB with a four-member deep ensemble, mean-variance estimation, and Monte Carlo dropout with four stochastic passes. The comparison is conducted on the aforementioned subset of 5,000 structures.
These four methods use different mechanisms to obtain an uncertainty score.
The ensemble estimates uncertainty from disagreement among four independently trained models; Monte Carlo dropout samples the stochastic predictor repeatedly with dropout enabled. Both MVE and EviDTB predict several parameters and calculate the uncertainty in one deterministic forward pass.
This distinction determines both the statistical comparison and the computational cost.
The model configurations are detailed in Sec.~S4 of the Supplementary Information.

During training (Fig.~\ref{fig:method_comparison}(g)), the ensemble costs approximately four times as much as the single-model methods. During inference, the ensemble and dropout each require four network evaluations, compared with one for MVE and EviDTB [Fig.~\ref{fig:method_comparison}(h)].

Figure~\ref{fig:method_comparison}(a)--(d) compares the mean uncertainty per atom with the mean local squared prediction error. EviDTB achieves Spearman-rank performance comparable to the ensemble method and higher than MVE and Monte Carlo dropout. It also attains the highest ROC-AUC for detecting high-error structures, outperforming even the ensemble [Fig.~\ref{fig:method_comparison}(e)]. In terms of Hamiltonian prediction accuracy, the ensemble yields the lowest errors, while EviDTB and MVE show comparable errors that are slightly higher than those of the ensemble [Fig.~\ref{fig:method_comparison}(f)].

Overall, EviDTB offers the best combined uncertainty-quantification performance among the compared methods, balancing error ranking and high-error detection with the low computational cost of a single forward pass.

\begin{figure*}[!tp]
  \includegraphics[width=\textwidth]{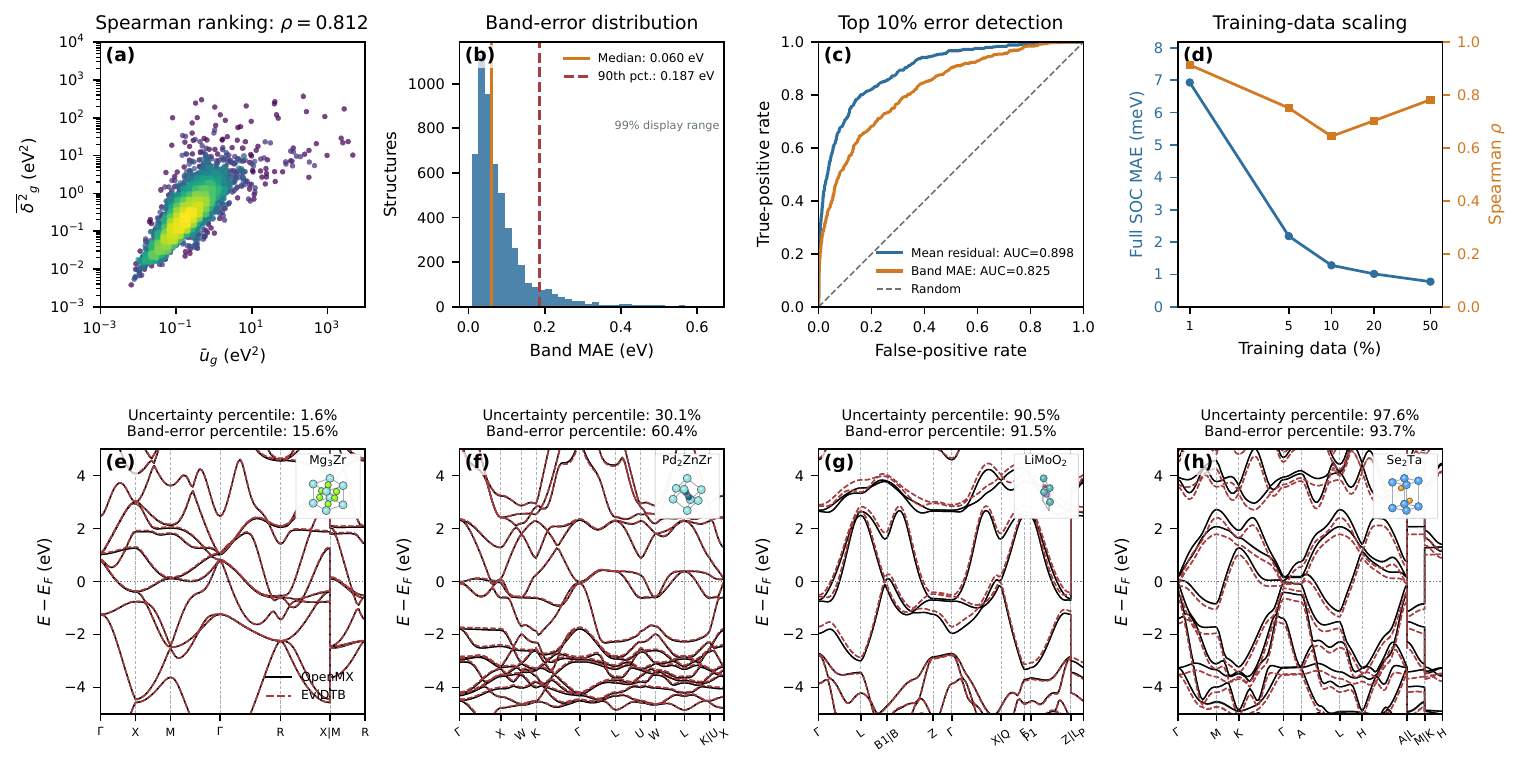}
    \caption{SOC evaluation on the 5,478 held-out test structures. (a) Mean uncertainty per atom, $\bar u_g$, versus mean local squared prediction error, $\overline{\delta^2}_g$, both in eV$^2$; Spearman $\rho=0.812$ includes structures outside the plotted limits. (b) Distribution of band MAE, with median \SI{0.060}{\electronvolt} and 90th percentile \SI{0.187}{\electronvolt}; the highest 1\% is omitted from the histogram only. (c) ROC curves using $\bar u_g$ to identify the highest-error 10\% by $\overline{\delta^2}_g$ (AUC 0.898) and by band MAE (AUC 0.825). (d) Full SOC matrix-element MAE and Spearman correlation between $\bar u_g$ and $\overline{\delta^2}_g$ for models trained on 1\%, 5\%, 10\%, 20\%, and 50\% of the training set, evaluated on the same test set. (e)--(h) OpenMX reference (black solid lines) and EviDTB-predicted (red dashed lines) bands for four selected test structures. The uncertainty/band-error percentile pairs are 1.6\%/15.6\%, 30.1\%/60.4\%, 90.5\%/91.5\%, and 97.6\%/93.7\%, respectively; insets show the corresponding crystal structures. The predicted bands use the spectral-projection method of UniTB~\cite{Liu2026UniTB} to suppress ghost-band artifacts. Band statistics and examples use the same overlap-filtered, spectrally clipped prediction and unprojected reference, as specified in Sec.~S2 of the Supplementary Information.}
  \label{fig:soc_evaluation}
\end{figure*}

\subsubsection{Evidential Deep Learning for Universal DFT Hamiltonian model}

Using the SOC data set described above, we trained an EviDTB model with approximately 17 million trainable parameters. This model achieves a matrix-element MAE of \SI{0.622}{\milli\electronvolt} over the reconstructed full complex SOC Hamiltonian elements of all 5,478 test structures.

Beyond average prediction accuracy, the evidential score provides a reliable structure-level ranking of Hamiltonian errors. On the held-out test set, the mean uncertainty per atom, $\bar u_g$, correlates with the mean local squared prediction error, $\overline{\delta^2}_g$, with Spearman $\rho=0.812$ [Fig.~\ref{fig:soc_evaluation}(a)]. Using $\bar u_g$ to identify the 10\% of structures with the largest $\overline{\delta^2}_g$ yields a ROC-AUC of 0.898 [Fig.~\ref{fig:soc_evaluation}(c)].

In practical applications, calculated or predicted Hamiltonians are often used to obtain electronic band structures. We therefore evaluate both band-prediction accuracy and the ability of the Hamiltonian uncertainty score to identify structures with large band errors.
To suppress ghost-band artifacts in the predicted bands, we adapt the spectral-projection method of UniTB~\cite{Liu2026UniTB}, as detailed in Sec.~S2 of the Supplementary Information.

\begin{table}[!htbp]
\caption{Structure-level prediction-error statistics before and after removing the 10\%, 15\%, or 20\% of structures with the highest mean uncertainty per atom, $\bar u_g$, on the held-out SOC test set. The mean local squared prediction error is $\overline{\delta^2}_g=N_g^{-1}\sum_{i\in g}\delta_i^2$. All statistics include the complete error tails. P95 and P99 denote the 95th and 99th percentiles. Parentheses give the percentage reduction relative to the unfiltered test set, calculated from unrounded statistics. Displayed error values are rounded to two decimal places, and reductions to the nearest integer percent.}
\label{tab:uncertainty_filtering}
\centering
\footnotesize
\setlength{\tabcolsep}{2.5pt}
\begin{tabular}{l r r r r}
\toprule
\shortstack[l]{Removed\\fraction} & \multicolumn{4}{c}{$\overline{\delta^2}_g$ (eV$^2$)} \\
\cmidrule(lr){2-5}
& Mean & Median & P95 & P99 \\
\midrule
0\% & 1.47 & 0.26 & 2.88 & 16.63 \\
10\% & 0.67 ($\downarrow$55\%) & 0.23 ($\downarrow$13\%) & 1.41 ($\downarrow$51\%) & 4.94 ($\downarrow$70\%) \\
15\% & 0.62 ($\downarrow$58\%) & 0.21 ($\downarrow$20\%) & 1.24 ($\downarrow$57\%) & 3.72 ($\downarrow$78\%) \\
20\% & 0.51 ($\downarrow$65\%) & 0.20 ($\downarrow$26\%) & 1.01 ($\downarrow$65\%) & 2.93 ($\downarrow$82\%) \\
\bottomrule
\end{tabular}
\end{table}

To quantify the benefit of uncertainty-guided screening, Table~\ref{tab:uncertainty_filtering} compares the full test set with the subsets retained after removing the 10\%, 15\%, and 20\% of structures with the highest mean uncertainty per atom, $\bar u_g$. For example, removing the highest 15\% reduces the mean local squared prediction error from 1.47 to 0.62~eV$^2$ and its 99th percentile from 16.63 to 3.72~eV$^2$, a reduction of 78\% in the latter. Thus, retaining approximately 85\% of the structures preferentially suppresses the high-error tail of the error distribution, without guaranteeing the accuracy of every retained prediction.

The band-error distribution has a long tail [Fig.~\ref{fig:soc_evaluation}(b)], with a median of \SI{0.060}{\electronvolt} and a 90th percentile of \SI{0.187}{\electronvolt}. For the 10\% largest band errors, the mean uncertainty per atom yields a ROC-AUC of 0.825 [Fig.~\ref{fig:soc_evaluation}(c)]. Thus, the score identifies high-error band predictions without further fine-tuning.

In the training-size comparison [Fig.~\ref{fig:soc_evaluation}(d)], the full SOC matrix-element MAE decreases from \SI{6.93}{\milli\electronvolt} at 1\% of the training set to \SI{0.78}{\milli\electronvolt} at 50\%, while the structure-level Spearman correlation ranges from 0.644 to 0.914 rather than increasing monotonically. Figure~\ref{fig:soc_evaluation}(e)--(h) compares predicted and reference bands for four selected test structures. Their uncertainty and band-error percentile ranks do not coincide exactly---for example, panel (f) is at the 30.1st uncertainty percentile but the 60.4th band-error percentile---so these examples illustrate individual predictions rather than establish an interval-averaged error trend.

\begin{figure}[!htbp]
  \centering
  \includegraphics[width=\columnwidth]{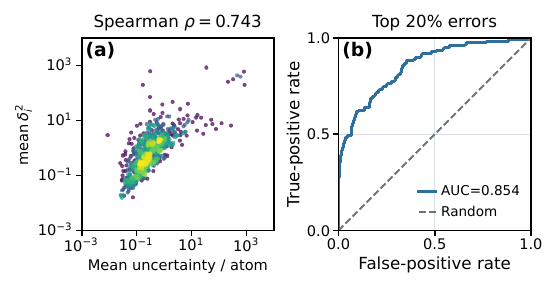}
   \caption{ Generated-structure evaluation on 636 structures without retraining. (a) Mean uncertainty per atom, $\bar u_g$, versus mean local squared prediction error, $\overline{\delta^2}_g$, both in eV$^2$ (Spearman $\rho=0.743$). (b) ROC curve using $\bar u_g$ to detect the highest-error 20\% of structures by $\overline{\delta^2}_g$ (AUC 0.854).}
  \label{fig:generated_structures}
\end{figure}

To further assess performance on an external data set, we evaluated crystal structures generated with CrystalFormer~\cite{Cao2025CrystalFormer}. We first excluded generated structures whose space group and chemical formula both matched an entry in the training, validation, or test set. The remaining structures were relaxed with MatterSim~\cite{Yang2024MatterSim} and used for OpenMX calculations~\cite{Ozaki2003OpenMX}; only structures for which both steps completed successfully were retained. The final external set contains 636 structures. Without retraining, the model achieves a full complex SOC Hamiltonian matrix-element MAE of \SI{1.095}{\milli\electronvolt}. The mean uncertainty per atom ranks the mean local squared prediction error with Spearman $\rho=0.743$ and detects the highest-error 20\% with ROC-AUC 0.854 [Fig.~\ref{fig:generated_structures}(a),(b)]. These results indicate that our model maintains good performance on new structures generated and relaxed using machine-learning models and can be incorporated into machine-learning-driven materials discovery workflows.

\subsection{Limitations}

The present results have several limitations. First, although deep evidential regression is commonly described as separating epistemic and aleatoric uncertainty, the validity of this interpretation remains under debate ~\cite{Amini2020,Meinert2023DERCritique}. We therefore evaluate only the performance of the combined uncertainty score, averaged per atom, and do not separately validate the epistemic and aleatoric components. Second, the crystal structures in this work are associated with the Materials Project ~\cite{Jain2013MaterialsProject}; the applicability of EviDTB to other screening scenarios, such as two-dimensional materials and high-pressure structures, requires further validation. Third, the DFT calculations assume that all structures are nonmagnetic. The present model is therefore restricted to nonmagnetic DFT Hamiltonians.

\section{Methods}

\subsection{Evidential regression formulation}

In the SLEM implementation of DeePTB, onsite blocks are stored on nodes and hopping blocks on directed edges. Each interaction has two edges, $i\to j$ and $j\to i$ (with opposite cell shifts for periodic images). On each node or edge, DeePTB stores orbital-shell pairs $a\leq b$ according to its fixed shell list, ordered first by angular momentum and then by radial-shell index. Each stored block contains all magnetic-quantum-number components; the complementary blocks are recovered by transposition. For equal-shell hopping blocks stored in both directions, the reconstructed Hamiltonian uses the average of the $i\to j$ block and the transpose of its $j\to i$ counterpart.

Let $\mathbf{y}_{ii}$ denote the valid stored onsite components on atom $i$, and $\mathbf{y}_{ij}$ those on an edge sourced at $i$. Their concatenation defines the local target $\mathbf{y}_i=(\mathbf{y}_{ii},\{\mathbf{y}_{ij}\}_{j:(i,j)\in\mathcal{E}_{+}})\in\mathbb{R}^{M_i}$. Here $\mathcal{E}_{+}$ is the graph's directed edge set, and the condition $(i,j)\in\mathcal{E}_{+}$ selects only outgoing edges with source $i$; incoming edges $j\to i$ do not enter $\mathbf{y}_i$. The dimension $M_i$ counts only valid stored components. We write the corresponding concatenated prediction as $\boldsymbol{\gamma}_i$.

In a localized atomic-orbital basis, the magnitudes of hopping Hamiltonian blocks decay approximately exponentially with interatomic distance $R$. Many matrix-element components consequently have small absolute amplitudes and are expected to contribute little to the local variance. Moreover, matrix elements are not statistically independent. Together, these properties motivate an effective statistical dimension of the local Hamiltonian distribution substantially smaller than the number of stored components. They motivate the low-dimensional approximation used below.

The unknown mean $\boldsymbol{\mu}_i$ has the same dimension $M_i$ as the local target and prediction. All distributions below are conditional on the local atomic environment. We approximate the local prediction error and fluctuations of the unknown mean around $\boldsymbol{\gamma}_i$ using a common isotropic $D$-dimensional subspace model. Let $U_i\in\mathbb{R}^{M_i\times D}$ have orthonormal columns spanning it, and define the covariance shape matrix

\begin{equation}
 P_i=U_iU_i^\top,
 \qquad U_i^\top U_i=I_D,
 \qquad \operatorname{tr}P_i=D.
 \label{eq:subspace_shape}
\end{equation}

Extending the scalar Gaussian--NIG construction of Ref.~\cite{Amini2020}, the corresponding hierarchy is

\begin{align}
 \mathbf{y}_i\mid\boldsymbol{\mu}_i,\sigma_i^2
 &\sim\mathcal{N}_{M_i}(\boldsymbol{\mu}_i,\sigma_i^2P_i),\nonumber\\
 \boldsymbol{\mu}_i\mid\sigma_i^2
 &\sim\mathcal{N}_{M_i}\!\left(\boldsymbol{\gamma}_i,\frac{\sigma_i^2}{\nu_i}P_i\right),\nonumber\\
 \sigma_i^2&\sim\mathrm{InvGamma}(\alpha_i,\beta_i).
 \label{eq:subspace_nig}
\end{align}

The singular Gaussian distributions are supported on the affine subspace $\boldsymbol{\gamma}_i+\operatorname{Range}(U_i)$, rather than on the entire ambient space. In orthonormal coordinates $\mathbf{z}_i=U_i^\top(\mathbf{y}_i-\boldsymbol{\gamma}_i)$ and $\mathbf{m}_i=U_i^\top(\boldsymbol{\mu}_i-\boldsymbol{\gamma}_i)$, this is an ordinary $D$-dimensional isotropic Gaussian--NIG model with $\mathbf{m}_i\mid\sigma_i^2\sim\mathcal{N}_D(\mathbf{0},\sigma_i^2I_D/\nu_i)$. Thus, the unknown mean retains its full Hamiltonian dimension; only its fluctuations are restricted to the effective subspace. Under the effective-subspace approximation, $\mathbf{y}_i-\boldsymbol{\gamma}_i\approx U_i\mathbf{z}_i$. We therefore approximate the full local squared residual by its effective-coordinate counterpart:

\begin{equation}
 \begin{aligned}
 \delta_i^2=\|\mathbf{y}_i-\boldsymbol{\gamma}_i\|_2^2
  &=\|\mathbf{y}_{ii}-\boldsymbol{\gamma}_{ii}\|_2^2
  +\sum_{j:\,(i,j)\in\mathcal{E}_{+}}
   \|\mathbf{y}_{ij}-\boldsymbol{\gamma}_{ij}\|_2^2\\
  &\approx\|\mathbf{z}_i\|_2^2.
 \end{aligned}
 \label{eq:pooled_error}
\end{equation}

Marginalizing over the unknown mean and variance yields a Student-$t$ distribution with $2\alpha_i$ degrees of freedom on this affine subspace. With $\Omega_i=2\beta_i(1+\nu_i)$, its density with respect to the subspace volume is

\begin{equation}
 p_i(\mathbf{y}_i)
 =\frac{\Gamma(\alpha_i+D/2)}{\Gamma(\alpha_i)}
 \left(\frac{\nu_i}{\pi\Omega_i}\right)^{D/2}
 \left(1+\frac{\nu_i\delta_i^2}{\Omega_i}\right)^{-(\alpha_i+D/2)}.
 \label{eq:subspace_predictive}
\end{equation}

Under this approximation, we use the full local squared residual in place of the effective-coordinate squared residual in the $D$-dimensional NLL, without explicitly constructing the subspace basis.

The effective dimension $D$ is used as a hyperparameter, and was set to 5 in this work.

With $\Omega_i=2\beta_i(1+\nu_i)$, taking the negative logarithm of Eq.~(\ref{eq:subspace_predictive}) and normalizing by $D$ gives the approximate predictive NLL used in training:
\begin{align}
 \mathcal{L}_{\mathrm{NLL},i}={}&\frac{1}{D}\bigg[\frac{D}{2}(\log\pi-\log\nu_i)
 -\alpha_i\log\Omega_i\nonumber\\
 &+\left(\alpha_i+\frac{D}{2}\right)
  \log(\nu_i\delta_i^2+\Omega_i)\nonumber\\
 &+\log\Gamma(\alpha_i)
  -\log\Gamma\!\left(\alpha_i+\frac{D}{2}\right)\bigg],
 \qquad D=5.
 \label{eq:nll}
\end{align}
This NLL is defined with respect to volume in the assumed $D$-dimensional affine subspace, not in the full ambient space. Its application to the full local residual vector uses the effective-subspace approximation described above.
The factor $1/D$ normalizes the NLL per effective dimension. The usual evidential penalty grows as $\delta_i(2\nu_i+\alpha_i)$, so a large initial prediction error can place excessive pressure on $\alpha_i$ and $\nu_i$. We instead saturate its error factor smoothly:
\begin{align}
 \mathcal{L}_{\mathrm{reg},i}
  &= (\alpha_i+2\nu_i)\tanh\!\left(\frac{|\delta_i|}{E_{H_0}}\right),\\
 \mathcal{L}_{\mathrm{real},i}
  &=\mathcal{L}_{\mathrm{NLL},i}+\lambda_r\mathcal{L}_{\mathrm{reg},i}+\lambda_e\delta_i^2,
 \label{eq:revised_loss}
\end{align}
where $E_{H_0}$ is the mean nodewise norm of the training residual target.
The hyperbolic tangent preserves a monotonic evidence penalty while preventing the initial large errors from dominating the evidence parameters.
The explicit squared-error term maintains direct pressure on Hamiltonian accuracy.
We use $\lambda_r=0.1$ and $\lambda_e=0.1$.

In the EviDTB implementation, raw node scalar outputs $a_i$ and $n_i$ give $\alpha_i=1+\operatorname{softplus}(a_i)+\epsilon$ and $\nu_i=\operatorname{softplus}(n_i)+\epsilon$, with $\epsilon=10^{-6}$. The scale $\beta_i$ is obtained from the raw node scale $b_i$ and outgoing-edge scales $b_{ij}$ as
\begin{equation}
 \beta_i=b_i^2+\sum_{j:\,(i,j)\in\mathcal{E}_{+}}b_{ij}^2.
 \label{eq:beta_pooling}
\end{equation}
These transformations ensure $\alpha_i>1$, $\nu_i>0$, and $\beta_i\geq0$. The Gaussian--NIG hierarchy gives the mean and covariance contributions
\begin{align}
 \mathbb{E}[\boldsymbol{\mu}_i]&=\boldsymbol{\gamma}_i,\nonumber\\
 \mathbb{E}[\sigma_i^2P_i]&=\frac{\beta_i}{\alpha_i-1}P_i,\nonumber\\
 \operatorname{Cov}(\boldsymbol{\mu}_i)
 &=\frac{\beta_i}{\nu_i(\alpha_i-1)}P_i.
 \label{eq:subspace_moments}
\end{align}
Within the effective-subspace model, adding the two covariance contributions and taking the trace gives the local total predictive variance,
\begin{equation}
 u_i=\operatorname{tr}\operatorname{Cov}(\mathbf{y}_i)
 =\frac{D\beta_i(1+\nu_i)}{\nu_i(\alpha_i-1)},
 \label{eq:vector_variance}
\end{equation}
which is evaluated directly from the predicted parameters without sampling and is used as an approximation to the expected squared local prediction error.

For a structure $g$ containing $N_g$ atoms, the error and uncertainty scores used in the ranking tests are their per-atom means,

\begin{equation}
 \overline{\delta^2}_g=\frac{1}{N_g}\sum_{i\in g}\delta_i^2,
 \qquad \bar u_g=\frac{1}{N_g}\sum_{i\in g}u_i.
 \label{eq:mean_structure_scores}
\end{equation}

\subsection{Projected reciprocal-space loss}

Real-space matrix-element accuracy does not by itself ensure accurate eigenvalues, particularly when the overlap matrix is ill-conditioned.
Following the projection strategy of NextHAM~\cite{Yin2025NextHAM}, we add a reciprocal-space objective.
For a sampled wave vector $\mathbf{k}$, the reference generalized eigensystem is
\begin{equation}
 H^{\mathrm{ref}}(\mathbf{k})U(\mathbf{k})
 =S(\mathbf{k})U(\mathbf{k})\varepsilon(\mathbf{k}).
\end{equation}
The predicted Hamiltonian is rotated into the reference eigenvector basis,
$\widetilde H=U^{\dagger}H^{\mathrm{pred}}U$, and partitioned into a target subspace $P$ and complementary subspace $Q$.
The $P$ subspace contains bands up to \SI{10}{\electronvolt} above the Fermi level.
The loss is
\begin{align}
 \mathcal{L}_{k}=w(t)\big[{}&
 \lambda_P\frac{\|\widetilde H_{PP}-\operatorname{diag}\varepsilon_P\|_F^2}{\max(1,N_P^2)}\nonumber\\
 &+\lambda_Q\frac{\|\widetilde H_{QQ}-\operatorname{diag}\varepsilon_Q\|_F^2}{\max(1,N_Q^2)}\nonumber\\
 &+\lambda_{PQ}\frac{\|\widetilde H_{PQ}\|_F^2}{\max(1,N_PN_Q)}\big],
 \label{eq:k_loss}
\end{align}
where $w(t)$ warms up linearly over 20 epochs. $N_P$ and $N_Q$ are the subspace sizes; empty blocks contribute zero. The block losses are mean squared errors in eV$^2$, averaged over sampled $\mathbf{k}$ points within each structure and then over structures.
We sample at most 16 $\mathbf{k}$ points per training structure and use
$(\lambda_P,\lambda_Q,\lambda_{PQ})=(2,5\times10^{-3},1.5)$.
The complete training objective is the node average of Eq.~(\ref{eq:revised_loss}) plus Eq.~(\ref{eq:k_loss}). The projected loss is inspired by the subspace-projection strategy of NextHAM; its implementation and weights are specific to EviDTB and are not a reproduction of the complete NextHAM training procedure.

\section{Conclusion}

We have introduced \EviDTB, which combines localized Hamiltonian learning with single-pass evidential uncertainty estimation. By learning a correction to a structure-dependent zeroth-step Hamiltonian, the model predicts nonmagnetic spin--orbit-coupled (SOC) Hamiltonians while providing an uncertainty score for each structure.

On 5,478 held-out structures, the reconstructed full SOC Hamiltonian achieves a matrix-element mean absolute error of \SI{0.622}{\milli\electronvolt}. The mean uncertainty per atom correlates with the mean local squared prediction error (Spearman $\rho=0.812$) and identifies the highest-error 10\% of structures with ROC-AUC 0.898. It also identifies the 10\% of structures with the largest band errors with ROC-AUC 0.825. Removing the 15\% most uncertain structures reduces the mean local squared prediction error from 1.47 to 0.62~eV$^2$ and its 99th percentile from 16.63 to 3.72~eV$^2$. On 636 generated structures evaluated without retraining, the model achieves a full SOC Hamiltonian matrix-element mean absolute error of \SI{1.095}{\milli\electronvolt}; uncertainty retains a Spearman correlation of 0.743 with the mean local squared prediction error and detects its highest-error 20\% with ROC-AUC 0.854.

These results show that the uncertainty score can help prioritize candidates for first-principles verification while Hamiltonian predictions rapidly cover a larger search space. We hope \EviDTB{} and uncertainty-guided screening will make high-throughput electronic-structure studies more efficient and accelerate the discovery of materials with desirable properties.

\begin{acknowledgments}
We thank Jiaheng Li for useful discussions. This work was supported by the Science Center of the National Natural Science Foundation of China (Grant No.~12188101), the National Key Research and Development Program of China (Grant No.~2023YFA1607400, 2022YFA1403800), the National Natural Science Foundation of China (Grant No.~12274436, 11921004), and H.W. acknowledge support from the New Cornerstone Science Foundation through the XPLORER PRIZE.
\end{acknowledgments}

\section*{Data and Code Availability}
The EviDTB code will be released as open-source software on GitHub shortly. A web demo is available for users to try EviDTB at \url{https://materialsgalaxy.iphy.ac.cn/EviDTB/}.

\bibliography{references}

\clearpage
\onecolumngrid
\begingroup
\def\EviDTBIncludeSupplementary{}
\def\EviDTBSupplementPath{supplementary/}
\ifdefined\EviDTBIncludeSupplementary
\setcounter{section}{0}
\setcounter{equation}{0}
\setcounter{table}{0}
\setcounter{figure}{0}
\renewcommand{\thesection}{S\arabic{section}}
\renewcommand{\theequation}{S\arabic{equation}}
\renewcommand{\thetable}{S\arabic{table}}
\renewcommand{\thefigure}{S\arabic{figure}}
\renewcommand{\theHsection}{supplement.\arabic{section}}
\renewcommand{\theHequation}{supplement.\arabic{equation}}
\renewcommand{\theHtable}{supplement.\arabic{table}}
\renewcommand{\theHfigure}{supplement.\arabic{figure}}
\begin{center}
\large\bfseries
Supplementary Information for:
Evidential Deep Regression for Universal Density Functional Theory Hamiltonian Models
\end{center}
\else
\documentclass[aps,prb,preprint,notitlepage,superscriptaddress,longbibliography]{revtex4-2}

\usepackage{amsmath,amssymb}
\usepackage{booktabs}
\usepackage{siunitx}
\usepackage{hyperref}
\sisetup{separate-uncertainty=true}
\newcommand{\EviDTB}{EviDTB}
\renewcommand{\thesection}{S\arabic{section}}
\renewcommand{\theequation}{S\arabic{equation}}
\renewcommand{\thetable}{S\arabic{table}}

\newcommand{\EviDTBSupplementPath}{}
\begin{document}

\title{Supplementary Information for:
Evidential Deep Regression for Universal Density Functional Theory Hamiltonian Models}

\author{Jiaxuan Liu}
\affiliation{Beijing National Laboratory for Condensed Matter Physics and Institute of Physics, Chinese Academy of Sciences, Beijing 100190, China}
\affiliation{University of Chinese Academy of Sciences, Beijing 100049, China}

\author{Yan Zhang}
\affiliation{Beijing National Laboratory for Condensed Matter Physics and Institute of Physics, Chinese Academy of Sciences, Beijing 100190, China}
\affiliation{University of Chinese Academy of Sciences, Beijing 100049, China}

\author{Caiyuan Ye}
\affiliation{Beijing National Laboratory for Condensed Matter Physics and Institute of Physics, Chinese Academy of Sciences, Beijing 100190, China}
\affiliation{University of Chinese Academy of Sciences, Beijing 100049, China}

\author{Zhong Fang}
\affiliation{Beijing National Laboratory for Condensed Matter Physics and Institute of Physics, Chinese Academy of Sciences, Beijing 100190, China}
\affiliation{University of Chinese Academy of Sciences, Beijing 100049, China}

\author{Hongming Weng}
\email{hmweng@iphy.ac.cn}
\affiliation{Beijing National Laboratory for Condensed Matter Physics and Institute of Physics, Chinese Academy of Sciences, Beijing 100190, China}
\affiliation{University of Chinese Academy of Sciences, Beijing 100049, China}

\author{Quansheng Wu}
\email{quansheng.wu@iphy.ac.cn}
\affiliation{Beijing National Laboratory for Condensed Matter Physics and Institute of Physics, Chinese Academy of Sciences, Beijing 100190, China}
\affiliation{University of Chinese Academy of Sciences, Beijing 100049, China}

\date{October 2026}
\maketitle
\fi

\section{Training settings}
\label{sec:training_settings}

We use the SLEM implementation in DeePTB as the representation backbone~\cite{Gu2024DeePTB,Zhouyin2025SLEM}.
Atoms are nodes and localized orbital interactions are directed edges within element-dependent cutoffs.
The model has four SLEM layers, spherical harmonics through $l=6$, and approximately 17 million trainable parameters.

The SOC training run uses Adam with an initial learning rate of $10^{-3}$ and a Reduce-on-Plateau scheduler monitored by the validation loss.
The batch size is five structures per rank, and training uses full float32 precision.
The reciprocal-space loss is warmed up over 20 epochs and samples at most 16 $\mathbf{k}$ points per training structure.
The reported SOC results use the checkpoint recorded at epoch 161.

\section{Band calculations and band-error statistics}
\label{sec:band_calculations}

We first Fourier transform the real-space Hamiltonian and overlap blocks to obtain $H_0^{\mathrm{SOC}}(\mathbf{k})$, $H_{\mathrm{ref}}^{\mathrm{SOC}}(\mathbf{k})$, $\Delta h^{\mathrm{pred}}(\mathbf{k})$, and $S(\mathbf{k})$, and then apply the following spectral-projection procedure at each sampled $\mathbf{k}$ point, adapting UniTB~\cite{Liu2026UniTB}.
The $\mathbf{k}$ dependence is suppressed below; $S$ denotes the full spinor overlap.
Retaining only overlap eigenmodes with $s_a>10^{-4}$ gives the canonical orthogonalization
\begin{equation}
 X=V_{\mathrm{k}}\operatorname{diag}(s_a^{-1/2})_{\mathrm{k}},
 \qquad X^{\dagger}SX=I,
 \label{eq:si_orthogonalization}
\end{equation}
where $V_{\mathrm{k}}$ contains the retained overlap eigenvectors.
Consistent with the main text, the predicted correction has two identical spin-diagonal blocks; in the orthonormal subspace,
\begin{equation}
 X^{\dagger}
 \begin{pmatrix}
  \Delta h^{\mathrm{pred}}&0\\
  0&\Delta h^{\mathrm{pred}}
 \end{pmatrix}X
 =W\operatorname{diag}(d_a)W^{\dagger}.
 \label{eq:si_correction_spectrum}
\end{equation}
We clip $d_a$ to $[-10,10]$ eV, obtaining $\widetilde d_a$, and compute the predicted bands as the eigenvalues of
\begin{equation}
 H_{\mathrm{band}}=X^{\dagger}H_0^{\mathrm{SOC}}X
 +W\operatorname{diag}(\widetilde d_a)W^{\dagger}.
 \label{eq:si_band_hamiltonian}
\end{equation}
These numerical cutoffs are specific to our implementation.
The reference bands solve the full generalized eigenproblem $H_{\mathrm{ref}}^{\mathrm{SOC}}c=\varepsilon^{\mathrm{ref}}Sc$, without overlap filtering or spectral clipping.
This band-only postprocessing does not change the reported Hamiltonian matrix-element errors or evidential uncertainty scores.

At each sampled $\mathbf{k}$ point, the predicted and reference energies are sorted in ascending order and paired by band index.
For each structure, band MAE is the mean of $|\varepsilon^{\mathrm{pred}}-\varepsilon^{\mathrm{ref}}|$ over all pairs whose reference energy satisfies $\varepsilon^{\mathrm{ref}}-E_F\leq\SI{10}{\electronvolt}$.
There is no lower energy cutoff or additional energy alignment, and every reference state in this window has a finite retained prediction partner.
The histogram, band-error ROC, percentile ranks, and band examples in Fig.~4 all use these same spectra and error values.
The displayed examples are restricted to $[-5,5]$ eV for readability; this display window does not define the band-MAE window.

Statistics use all 5,478 held-out structures.
The median and 90th-percentile band MAEs are \SI{0.060}{\electronvolt} and \SI{0.187}{\electronvolt}, respectively.
The histogram alone omits the highest-error 1\% for display.
The band-error ROC uses the mean uncertainty per atom to identify structures at or above the 90th percentile of band MAE, giving ROC-AUC 0.825.
Both uncertainty and band-error percentile ranks are computed as $100(r-1)/(5478-1)$, with average ranks for ties.
The four illustrated materials are fixed examples, not averages over percentile bins.

\section{Spin-block structure of the Hamiltonian correction}
\label{sec:spin_blocks}

Following the spin-block analysis of Ref.~\cite{Yin2025NextHAM}, we evaluate the Frobenius norms of the real and imaginary parts of each SOC correction block $H-H_0$ for each material.
Table~\ref{tab:si_soc_block_max} gives their maxima over the 54,895 structures in the data set.
The spin off-diagonal and imaginary corrections are negligible compared with the real spin-diagonal corrections, supporting the real spin-diagonal correction learned by \EviDTB.
The remaining SOC contributions are retained from $H_0^{\mathrm{SOC}}$.

\begin{table}[htbp]
\caption{Maximum per-material Frobenius norms of the real and imaginary parts of the SOC Hamiltonian correction $H-H_0$, over 54,895 materials. Norms are in eV.}
\label{tab:si_soc_block_max}
\begin{ruledtabular}
\begin{tabular}{lcccc}
Spin block & $\uparrow\uparrow$ & $\uparrow\downarrow$ & $\downarrow\uparrow$ & $\downarrow\downarrow$ \\
Real part & $237.25$ & $<10^{-6}$ & $<10^{-6}$ & $237.25$ \\
Imaginary part & $<10^{-6}$ & $<10^{-6}$ & $<10^{-6}$ & $<10^{-6}$ \\
\end{tabular}
\end{ruledtabular}
\end{table}

\section{Configurations of the uncertainty-estimation models}
\label{sec:uq_models}

The four methods compared in Fig.~3 use the same 4,000 training, 500 validation, and 500 test structures.
Their shared DeePTB/SLEM backbone has four layers, angular momenta through $l=6$, 32 radial basis functions, and hidden irreps
$128\times0e+64\times1o+64\times2e+32\times3o+32\times4e+16\times5o+8\times6e$.
Each model is trained for 130 epochs using Adam with an initial learning rate of $10^{-3}$, Reduce-on-Plateau scheduling, float32 precision, and one structure per GPU on four GPUs.
The single-model methods use random seed 20260401.

All methods include a direct real-space squared-error term, $\mathcal{L}_{\mathrm{SE}}=\langle\delta_i^2\rangle_i$, where the average runs over valid source nodes, and the projected reciprocal-space loss defined in Methods.
For this small-subset comparison, the reciprocal-space weights are $(\lambda_P,\lambda_Q,\lambda_{PQ})=(2\times10^{-3},5\times10^{-6},1.5\times10^{-3})$, with a 20-epoch linear warmup and at most 16 randomly sampled $\mathbf{k}$ points per training structure.
We denote this loss by $\mathcal{L}_{k}^{\mathrm{subset}}$ to distinguish its weights from those of the full-data model.

\subsection{Deep ensemble}

We independently train four complete DeePTB/SLEM models using four different random seeds and the same training and validation structures.
Each member minimizes $0.5\mathcal{L}_{\mathrm{SE}}+\mathcal{L}_{k}^{\mathrm{subset}}$.
The ensemble prediction is the mean of the four Hamiltonian predictions, and their dispersion supplies the uncertainty estimate.

\subsection{Monte Carlo dropout}

We train one model with dropout probability $p=0.1$ on the equivariant feature routes after each SO(2)-based SLEM layer.
Dropout acts on complete irreducible-representation copies: all magnetic-quantum-number components of a copy share the same Bernoulli mask and are scaled by $1/(1-p)$ when retained.
The mask therefore commutes with rotations and preserves equivariance for a fixed stochastic realization.
Dropout is applied to node, edge, and hidden features, with the two directions of an edge sharing masks within each edge-feature branch.
Training minimizes $0.5\mathcal{L}_{\mathrm{SE}}+\mathcal{L}_{k}^{\mathrm{subset}}$.
At prediction time, dropout remains enabled for four stochastic passes; their mean gives the Hamiltonian prediction and their dispersion gives the uncertainty.

\subsection{Mean-variance estimation}

The MVE model predicts the Hamiltonian mean and one scalar variance for each source-centered local target.
Scalar node and outgoing-edge outputs $b_i$ and $b_{ij}$ give
\begin{equation}
 v_i=b_i^2+\sum_{j:\,(i,j)\in\mathcal{E}_{+}}b_{ij}^2.
 \label{eq:si_mve_variance}
\end{equation}
Using the same effective-subspace assumption as in the main text, we model the signed local error coordinates as $\mathbf{z}_i\sim\mathcal{N}_D(\mathbf{0},v_i I_D)$, with $D=5$ and $\|\mathbf{z}_i\|_2^2=\delta_i^2$.
The Gaussian density and its negative logarithm per effective dimension are therefore
\begin{align}
 p(\mathbf{z}_i)&=(2\pi v_i)^{-D/2}
 \exp\!\left(-\frac{\delta_i^2}{2v_i}\right),
 \label{eq:si_mve_density}\\
 \mathcal{L}_{\mathrm{NLL},i}^{\mathrm{MVE}}
 &=-\frac{1}{D}\log p(\mathbf{z}_i)
 =\frac{1}{2}\left[\log(2\pi v_i)+\frac{\delta_i^2}{Dv_i}\right].
 \label{eq:si_mve_nll}
\end{align}
The variance is floored at $10^{-6}$ when evaluating the NLL.
The complete objective is
\begin{equation}
 \mathcal{L}_{\mathrm{MVE}}
 =0.5\mathcal{L}_{\mathrm{SE}}
 +w_{\mathrm{NLL}}\bigl\langle\mathcal{L}_{\mathrm{NLL},i}^{\mathrm{MVE}}\bigr\rangle_i
 +\mathcal{L}_{k}^{\mathrm{subset}},
 \label{eq:si_mve_loss}
\end{equation}
where $w_{\mathrm{NLL}}=0$ for the first 20 epochs and $0.1$ thereafter.
The total local predictive variance is $u_i^{\mathrm{MVE}}=Dv_i$, available in one forward pass.

\subsection{EviDTB}

EviDTB uses the Gaussian--NIG model and bounded evidence regularizer derived in Methods, with $D=5$ and $\lambda_r=0.1$.
For the small-subset comparison, the direct squared-error weight is $\lambda_e=0.5$, and the reciprocal-space term is $\mathcal{L}_{k}^{\mathrm{subset}}$.
The Hamiltonian and total local predictive variance are obtained in one forward pass.
For all four methods, structure-level scores are averaged over atoms as defined in the main text.

\ifdefined\EviDTBIncludeSupplementary
\else
\bibliography{references}
\fi

\ifdefined\EviDTBIncludeSupplementary\expandafter \fi

\clearpage
\endgroup

\end{document}